\documentclass[a4paper]{article}
\pdfoutput=1 

\usepackage{jcappub}  
\usepackage{float}
\usepackage[T1]{fontenc} 
\usepackage[english]{babel}
\usepackage{graphicx}
\usepackage{siunitx}
\usepackage{amsmath,latexsym}
\usepackage[version=3]{mhchem}
\usepackage{amssymb}
\usepackage{subfig}

\usepackage{xcolor}
\usepackage{multirow}
\usepackage{enumerate}
\usepackage[symbol]{footmisc}
\usepackage{awesomebox}
\usepackage[section]{placeins}

\DeclareSIUnit \belm {Bm}
\DeclareSIUnit{\Sample}{S}
\DeclareSIUnit\year{yr}
\usepackage[textwidth=2.8cm]{todonotes}

\title{First results from BRASS-p broadband searches for hidden photon dark matter}

\author[a]{Fayez Bajjali,}
\author[b]{Sven Dornbusch,}
\author[a]{Marko Ekmedžić,}
\author[a]{Dieter Horns,}
\author[b]{Christoph Kasemann,}
\author[b]{Andrei Lobanov,}
\author[a]{Artak Mkrtchyan,}
\author[a]{Le Hoang Nguyen,}
\author[a]{Martin Tluczykont,}
\author[b,c]{Gino Tuccari,}
\author[a]{Johannes Ulrichs,}
\author[b]{Gundolf Wieching,}
\author[b]{Anton Zensus}
\emailAdd{le.hoang.nguyen@uni-hamburg.de, alobanov@mpifr-bonn.mpg.de, dieter.horns@uni-hamburg.de}

\affiliation[a]{Institut f{\"u}r Experimentalphysik, Universit{\"a}t Hamburg, Hamburg, Germany.}
\affiliation[b]{Max-Planck-Institut f{\"u}r Radioastronomie, Bonn, Germany.}
\affiliation[c]{INAF Istituto di Radioastronomia, Bologna, Italy}

 \abstract{
We discuss first results from hidden photon dark matter searches made with a prototype of the Broadband Radiometric Axion/ALPs Search Setup (BRASS-p) in the range of particle mass of \SIrange{49.63}{74.44}{\micro\electronvolt} (frequency range of \SIrange{12}{18}{\giga\hertz}). The conceptual design of BRASS and a detailed description of its present prototype, BRASS-p, are given, with a view of the potential application of such setups to hidden photon, axion, and axion-like particle (ALP) dark matter searches using heterodyne detectors in the range of particle mass from \SI{40}{\micro\electronvolt} to \SI{4000}{\micro\electronvolt} (\SI{10}{\giga\hertz} to \SI{1}{\tera\hertz}). Pioneering measurements made with BRASS-p achieve the record sensitivity of (0.3--1.0)$\times$$10^{-13}$ to the kinetic mixing between the normal and hidden photons, assuming the dark matter is made entirely of unpolarized hidden photons. Based on these results, a discussion of further prospects for dark matter searches using the BRASS-p apparatus is presented.
}

\begin{document}
\maketitle

\section{Introduction}

Many well-motivated extensions of the  Standard Model (SM) of Particle Physics contain 
a hidden (or dark) sector which provides an explanation for the missing components observed in the Universe while remaining minimally coupled via gravity to the common forms of matter (see \citep{Battaglieri2017} and references therein). 
One of such missing, invisible components in the Universe is 
the so-called dark matter that is responsible for the largest fraction, 
85 \%, of the mass content of the Universe \citep{2020A&A...641A...6P}. 
The best-studied dark matter particle candidate is a weakly interacting, 
massive particle (WIMP), that was produced thermally
during the early Universe and is naturally predicted, for instance, as the lightest 
supersymmetric particle (LSP) in widely-discussed extensions of the SM \citep{1996PhR...267..195J}.

So far,  direct and indirect searches for WIMPs have not revealed the presence 
of WIMP-type dark matter. Further improvements in these searches approach 
some intrinsic limitations. Direct searches for elastic scattering of WIMPs in 
well-shielded underground laboratories are eventually limited by
the \textit{neutrino fog} of  
background events from solar and cosmogenic (core-collapse supernova) neutrinos \cite{Akerib:2022ort}. 
Indirect searches (e.g., gamma-ray emission from self-annihilating
LSPs) are limited by the presence of cosmic accelerators that produce
an indistinguishable gamma-ray background. 
Collider-based searches are limited to searches for particles that are not
predicted in the SM. Even though this is not to be confused with searching for
dark matter, so far no conclusive evidence for deviations from the SM have been found. 

Presently, the notion of WIMP-type dark matter is theoretically well-established, 
but there are so far neither experimentally nor observationally established
indications for the existence of WIMPs. 
Among the non-WIMP particle-type candidates for dark matter, 
Weakly Interacting Slim Particles (WISPs) have gained increased 
attention in the past years. The two main representatives of WISPs 
are the axion or axion-like particles (ALPSs) and hidden photons (HPs) 
\cite{essigDarkSectorsNew2013,ariasColdDarkMatter2012,ringwaldAxionsAxionlikeParticles2014,2020PhR...870....1D}. 
Even though the underlying motivations for the introduction of axions/ALPs and HPs and their respective cosmological phenomenology 
are very different, the search strategies in laboratory experiments 
share some similarities \cite{sikivie_experimental_1984}. 
An experiment that searches for WISPs particle sourced from the halo dark matter is called \textit{a haloscope}
Almost every haloscope experiment searching for WISPs relies on a photon signal that originates from the conversion of WISPs to photons. In the case of pseudo-scalar axion/ALPs, the inverse Primakoff effect in the presence of an external magnetic field allows the conversion from WISPs to photon. 
For HPs, the relevant process is the kinetic mixing between hidden photons and photons \cite{ariasColdDarkMatter2012,abelKineticMixingPhoton2008}. 

The  resulting signal of the converted photons is often enhanced using a low dissipation resonant cavity e.g., ADMX 
\cite{duSearchInvisibleAxion2018}, WISPDMX \cite{Nguyen:2019xuh},
HAYSTAC \cite{brubakerFirstResultsMicrowave2017,backesQuantumEnhancedSearch2021}, ORGAN \cite{McAllister:2022ibe}, and various cavity-based experiments in CAPP, Korea \cite{semertzidisAxionDarkMatter2019,kwonFirstResultsAxion2021}. Cavity-based 
haloscopes that are operated at a resonant frequency $\nu_{cav}$ 
are limited to a WISP-mass $m$ such that $h\nu_{cav} \approx mc^2$, neglecting
the kinetic energy of the gravitationally bound dark matter halo. The linear
dimension of the resonating cavity scales approximately with the Compton wavelength of the dark matter particle,  
\begin{equation}
\lambda \approx  6~\mathrm{cm} \left(\frac{mc^2}{20~\mu\mathrm{eV}}\right)^{-1},
\end{equation}
and therefore the volume of the cavity becomes progressively smaller for higher particle masses. 
For the mass range beyond tens of  $\mu$eV
and up to a few \textit{m}eV, 
experiments with resonant enhancement from metamaterials \cite{lawsonTunableAxionPlasma2019,schutte-engelAxionQuasiparticlesAxion2021} or boosting from dielectric material  stacks \cite{brunNewExperimentalApproach2019,majorovitsMADMAXNewRoad2020} are under development.

A different approach to search for DM makes use of the conversion of HPs 
into an electromagnetic wave at the surface of a conductor 
\cite{hornsSearchingWISPyCold2013}. For HP with a mass, $m$, and a photon coupling, $\chi$, the conversion results from excitation of free electrons on the conductor surface by the residual electric field $\textbf{E}_\mathrm{hp} = \chi \, m\, \textbf{X}_\mathrm{hp}$ vested in the HP dark matter field, $\textbf{X}_\mathrm{hp}$, propagating through the conductor. For conversion of an axion/ALPs with a mass, $m$, the respective electric field is induced by a transverse magnetic field, $B_\mathrm{\perp}$, (the field component oriented parallel to the conducting surface)extending up to at least a wavelength, $\lambda_\mathrm{a} = 2\pi/m$, of the electromagnetic wave generated by the conversion. With such settings, the measurement sensitivity to the HP dark matter signal is proportional to the area $A_\mathrm{conv}$ of the conversion surface, while for the axion/ALPs dark matter it is $\propto A_\mathrm{conv} B_\mathrm{\perp}^2$.

Adopting this approach, we describe here the first prototype of a Broadband Radiometric axion/ALPs Search Setup (BRASS) which will ultimately aim to employ a permanently magnetised conversion surface with $A_\mathrm{conv} \approx 100~\mathrm{m}^2$ and $B_\mathrm{\perp} \approx 1$~T. The BRASS prototype (BRASS-p) presently operates with $A_\mathrm{conv} = 6.0~\mathrm{m}^2$ and does not have the permanent magnet panels in place, which limits its current application to searches for HP dark matter. The physical foundations and the conceptual design of BRASS-p measurements are described in Section~\ref{Section:HPSignal}. The current BRASS-p setup is presented and discussed in Section~\ref{Section:Setup}. Section~\ref{Section:Results} describes the system noise calibration procedures, and the data taking run from
the HP dark matter search with BRASS-p in the extended frequency-range
from $\nu=12~\mathrm{GHz}$ to $\nu=18~\mathrm{GHz}$,
which corresponds to the HP mass range 
from $m=41.35~\mu\mathrm{eV}$ to 
$m=74.44~\mu\mathrm{eV}$. 

\section{Hidden photon dark matter signal in BRASS}
\label{Section:HPSignal}
\subsection{Hidden photon conversion}
 The conversion of HP dark matter to detectable photons
 is driven by the kinetic mixing \cite{HOLDOM1986196,FOOT199167}
 of the HP $\tilde X^\mu$ (field $\tilde X^{\mu\nu}$) 
 with photons $A^\mu$ (field $F^{\mu\nu}$). The strength of 
 the mixing is related to the mixing angle $\chi\ll 1$ that is 
 predicted to typical values between $\chi\approx 10^{-12}$
 and $\chi\approx 10^{-3}$ (see reference \cite{hornsSearchingWISPyCold2013} and references therein).
  
 The  
 electric field that results from the kinetic mixing is given by 
 \begin{equation}
 \label{eqn:hpmix}
     \mathbf{E}_\text{hp} =  \chi m \mathbf{X}_\text{hp},
 \end{equation}
 with $\sqrt{|\mathbf{X}_\text{hp}|^2}=m\sqrt{2\rho_{DM}}$
 for a given local dark matter density $\rho_{DM}$.\footnote{The natural units with $c=\hbar=1$ are assumed throughout the text.}
 
 The resulting (time-averaged) amplitude of the electric field is
 \begin{equation}
 \label{eq:hp_fieldstrength}
    \sqrt{\langle |\mathbf{E}_\text{hp}|^2\rangle} = 
    \chi \sqrt{2\rho_{DM}} \approx 3.3\times 10^{-9} \frac{\mathrm{V}}{\text{m}} \left(\frac{\chi}{10^{-12}}\right)
    \left( 
    \frac{\rho_{DM}}{0.3~\mathrm{GeV}/\mathrm{cm}^3}
    \right).
 \end{equation}

 At the boundary of a conductor
 surface,  the tangential component of the  electric 
 field $\bf{E}_\text{hp}$ that results from the mixing with the HP field  
 oscillates at the frequency $\nu\approx 4.8~\mathrm{GHz} (m/20~\mu\mathrm{eV})$.
  The oscillating electrons in the conductor will emit radiation perpendicular to the conductor surface \cite{hornsSearchingWISPyCold2013}. 

The  strength of the field that is relevant for conversion 
is therefore the 
component tangential $\mathbf{E}_{hp,\parallel}$ to the 
 surface of the conductor. This component  
 is maximal for a polarisation
 of $\bf{X}_{hp}$ parallel to the surface ($\vartheta=0$) and
 disappears for a perpendicular polarisation ($\vartheta=\pi/2$). 
 In general, $|\mathbf{E}_{hp,\parallel}|=\chi m |\cos\vartheta| |\mathbf{X}_{hp}|$
 depends on the orientation of the polarisation of the
 HP $\bf{X}_{hp}$.  
 
 In the following, we consider the scenario in which 
 the hidden photon dark matter particles behave like a gas, and 
 their polarisation direction points to random directions 
 everywhere in space. Hence, the averaged emitted power $P_\text{hp}$ over the 
 surface area of the conversion surface $A$ is given by 
  \begin{equation}
     P_\text{hp} = A\langle |\mathbf{E}_{hp,\parallel}|^2\rangle = A\frac{2}{3} 
     \chi^2 \rho_{DM}\, . \label{eq:power_dish}
 \end{equation}
 In DM search experiments, this expectation is matched against the actual output power $P_\mathrm{det}$ or (in case of non-detection) the standard deviation sensitivity (noise power) $\sigma_\mathrm{det}$ of the measurement, thereby constraining the HP coupling strength, $\chi$. In experiments in which individual photons cannot be counted, the measured power and sensitivity of detection are described most generally by the radiometer equation
 \begin{equation}
     P_\mathrm{det} = (\mathrm{S/N})\, \sigma_\mathrm{det} = (\mathrm{S/N})\, k_\mathrm{B} T_\mathrm{sys}\, \mathrm{BW}^{1/2}\, \tau^{-1/2} \,,
     \label{eq:radiometer}
 \end{equation}
 where S/N is the signal-to-noise level of detection, or exclusion, of the signal, $k_\mathrm{B}$ is the Boltzmann constant, $T_\mathrm{sys}$ is the system noise temperature of the measurement apparatus, and BW and $\tau$ are the measurement bandwidth and duration. 
 
 It is worthwhile mentioning, that the expected 
 signal is at all times fully linearly polarised \citep{caputoDarkPhotonLimits2021}.
Intriguing enough, the direction of the polarisation can change
smoothly over time for the case where the orientation
of $\bf{X}_{hp}$ remains constant over some extended 
region in space such that
the local polarisation direction in the rotating laboratory frame changes smoothly and in a predictable way with time.
However, we are not considering these more model-dependent scenarios here and leave it for future investigations and future data-taking.

For the DM material located in a virialised Galactic halo, the received signal is expected to have a Maxwellian profile with a characteristic fractional width of ${\Delta \nu}/{\nu} \sim 10^{-6}$. For non-virialised scenarios involving DM streams, signals with ${\Delta \nu}/{\nu} \sim 10^{-7}$ or even smaller can be expected \cite{OHare:2017yze,Chakrabarty:2020qgm}. To optimise the detection probability of such signals, measurements with the fractional spectral resolution down to $\sim 10^{-9}$ may be needed.

\subsection{Conceptual design of BRASS}
\label{sec:optical_scheme}

\begin{figure}[!ht]
    \centering
    \setlength{\unitlength}{0.05\textwidth}
    \begin{picture}(14,7)
        \put(3.5,0){\includegraphics[scale =0.3]{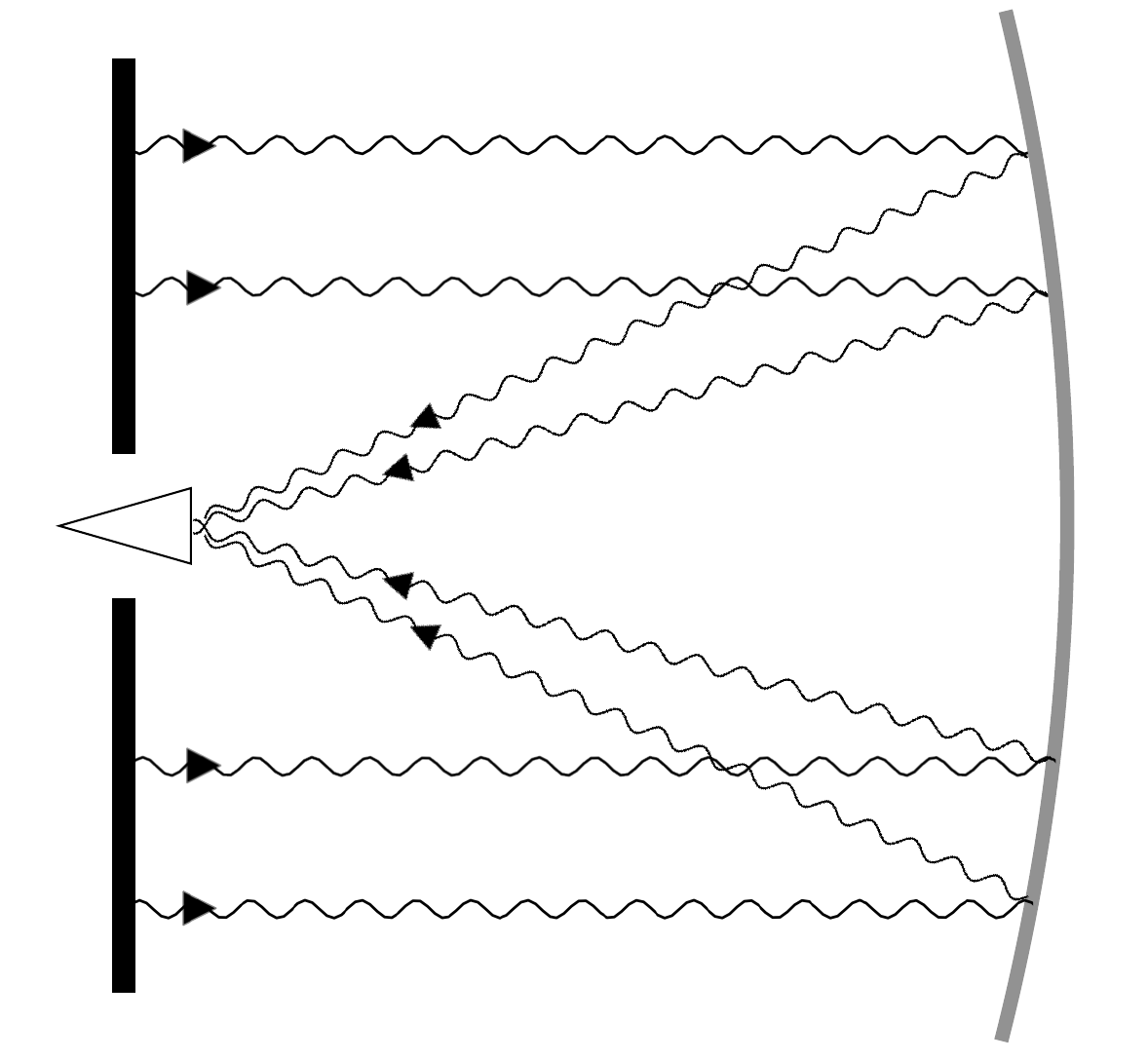}}
        \put(3.6,6.5){$\mathcal{A}$}

        \put(3.3,3.6){$\mathcal{D}$}
        \put(5,5.8){$\gamma_\text{hp}$}
        \put(5,1.4){$\gamma_\text{hp}$}
    \end{picture}
    \caption{
    Conceptual optical scheme of BRASS. A flat, conducting, and permanently magnetised conversion surface ($\mathcal{A}$) is employed for generating an electromagnetic signal arising in response to the electric field of the dark matter propagating through. This electric field is either inherent to the DM particles themselves (for hidden photons) or induced by the magnetic field near the conversion surface (for axions/ALPs). The resulting electromagnetic signal is focused by the parabolic reflector and detected using a heterodyne receiver ($\mathcal{D}$) with a broadband digital backend.
    }
    \label{fig:BRASSDesign}
\end{figure}
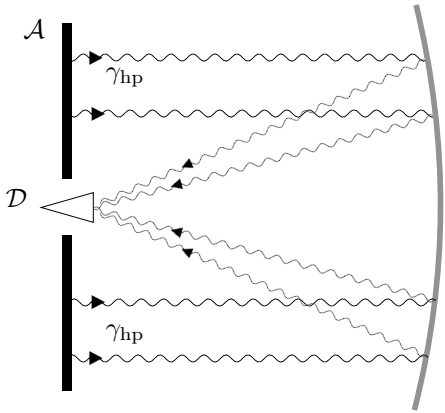

The conceptual optical design of BRASS (Fig.~\ref{fig:BRASSDesign}) comprises a shielded chamber with a flat, conducting {\em conversion surface} at which the electromagnetic signal from HP or axion/ALPs DM is generated and a parabolic mirror concentrating this signal onto a detector. A similar setup employing commercial parabolic antennas has also been discussed \cite{Suzuki:2015vka}. Operations with multiple chambers are also principally feasible, allowing for enhancing the detection and directional sensitivity. The actual setup of the BRASS prototype experiment operates with a single chamber and employs a $D=2.5$~m reflector with a focal distance of $f=4.8$~m. 


The outgoing electromagnetic wave generated at the conversion surface by the passage of DM particles can be treated as a plane wave if the electric field $\mathbf{E}_\mathrm{hp}$ 
oscillates in phase over the full area of the conversion surface. This conditions is satisfied
for as long as the coherence length of the HP field is large against the linear dimension of
the conversion surface. The coherence length can be approximated by the de Broglie wavelength, 
\begin{equation}
    \lambda_{d.B.} \approx \frac{2\pi}{m\,\beta} \approx 60~\mathrm{m} \frac{20~\mu\mathrm{eV}}{m} \frac{10^{-3}}{\beta}\,,
\end{equation}
where $\beta\approx 10^{-3}$ the velocity of the dark matter flow.
For the conversion surface of BRASS-p with a linear dimension
$d=2.5~\mathrm{m}$, the oscillations will be quasi-coherent for masses  
$m\lesssim 0.5~\mathrm{meV}$. 

To quantify the signal properties, taking into account the diffraction and reflections within the apparatus, we simulate the optical setup of BRASS-p using the finite element method (FEM) implemented in the COMSOL Multiphysics$^\text{®}$ package.
We adopt the spatial geometry of BRASS and simplify the problem to two dimensions, so 
that the essential components (the conversion surface and the
parabolic reflector) with their respective geometry are embedded through matching boundary conditions that mimic an infinite free space. 
The simulation setup is constructed assuming that the electric field ($ \mathbf{E}_\mathrm{hp}$) is generated at the boundary of the conversion surface and it excites a plane wave propagating outward. The surface of the parabolic mirror is simulated as a perfect electric conductor. The entire setup is spatially enclosed by a perfect absorbing surface which approximates the free space conditions.

\begin{figure}[!ht]
    \centering
    \setlength{\unitlength}{0.05\textwidth}
    \begin{picture}(14,12.5)
        \put(-3,0.5){\includegraphics[scale =0.25]{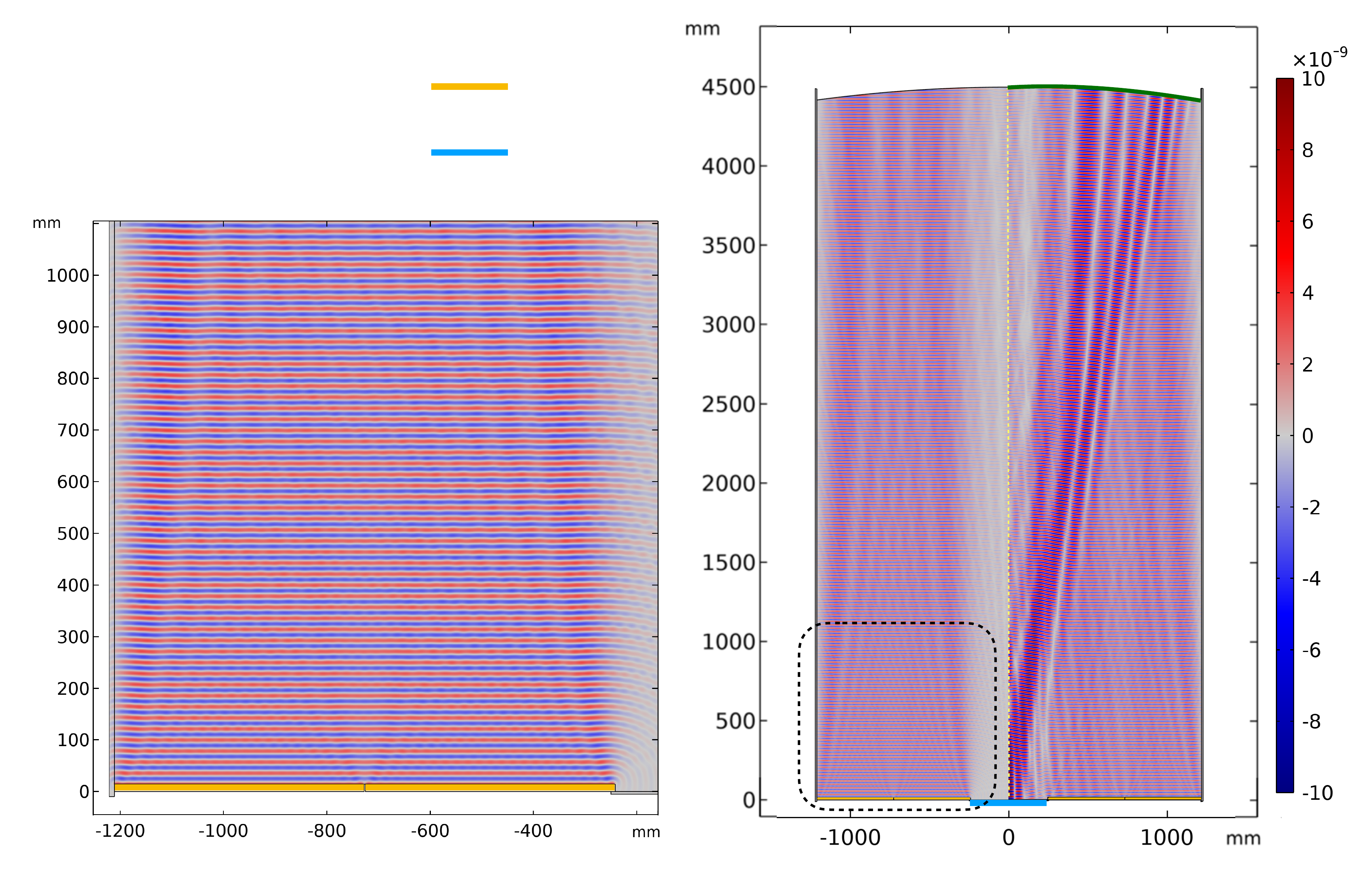}}
  
        \put(2,0){(a)}
        \put(10.9,0){(b)}
        \put(16,4.5){\rotatebox{90}{\small{Electric Field (V/m)}}}
        \put(11.8,11.8){\rotatebox{-3}{\small{PEC}}}
        \put(-1,11.5){{\small{Conducting surface}}}
        \put(1,10.55){{\small{Receiver}}}
    \end{picture}
    \caption{HP signal propagation. The coupling factor $\chi= 10^{-13}$ is selected for the purpose of demonstration. (a) The $z$ component of the electric field is excited by the conversion of hidden photon dark matter at the surface of the conductor. (b) This compound figure shows the propagation of the plane waves in two scenarios: with (left) and without (right) the parabolic mirror (the green edge at the top right). The dashed box indicates the region presented by (a).}
    \label{fig:fem_simulation}
\end{figure}
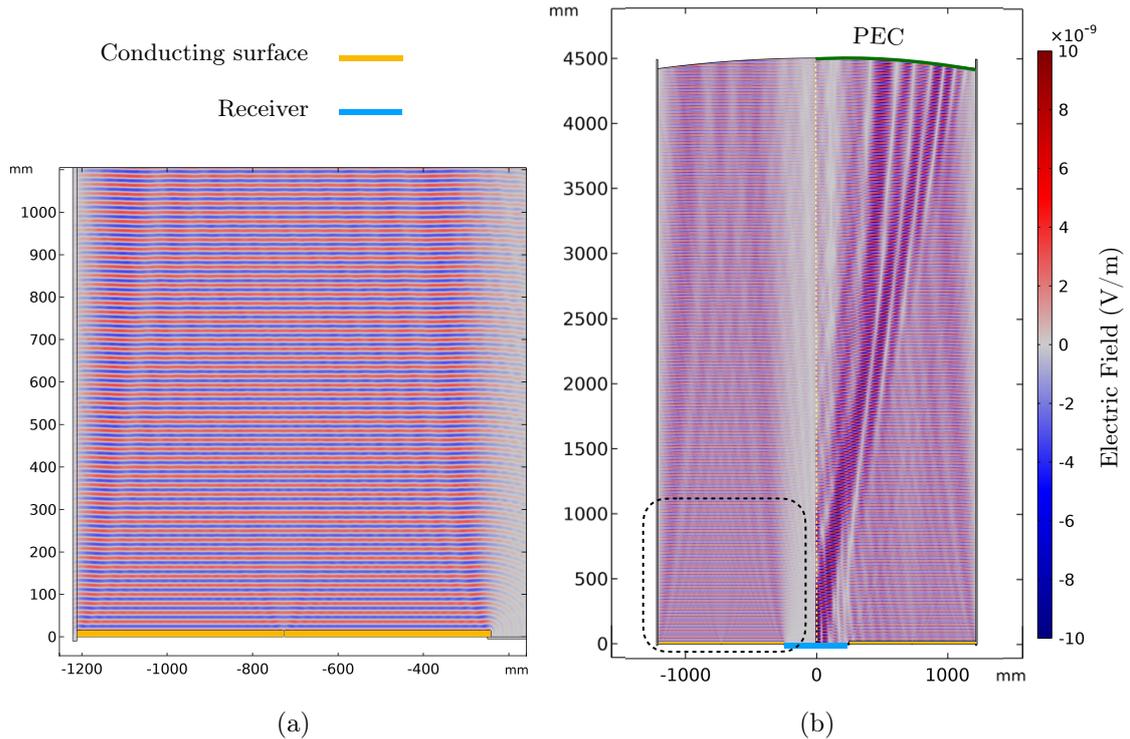

The simulation is performed for a frequency of 14~GHz, and the magnitude of the simulated electric field corresponds to the signal strength expected from the hidden photon dark matter with the photon coupling $\chi=10^{-13}$. The maximum field strength obtained in the simulation is $\approx 5\times 10^{-10}~\mathrm{V/m}$, which is consistent with the expected time averaged field strength given by Eqn.~\ref{eq:hp_fieldstrength}. 

An example of the resulting instantaneous
electromagnetic field near the conversion surface is shown in Fig.~\ref{fig:fem_simulation}a, where the diffraction at the edges of the surface can be clearly seen. The effect of the parabolic reflector on the electric field distribution can be assessed from  Fig.~\ref{fig:fem_simulation}b, which compares the generated field freely propagating outward  (left panel), to the one focused by the parabolic mirror onto the detector (right panel).

The overall optical efficiency, $\eta_\mathrm{opt}$, of the signal focusing can be estimated by integrating the total power in the freely outgoing electromagnetic wave to the total power at the surface of the receiver aperture. For the simulations shown in Fig.~\ref{fig:fem_simulation}, $\eta_\mathrm{opt} \approx 0.98$ is obtained. The simulation shows a good approximation of the BRASS-p optical efficiency in the entire 12 GHz to 18 GHz frequency range for which $\lambda_\mathrm{d.B.}/D = 6.4$--$11.6$.

\section{BRASS-p}
\label{Section:Setup}

\begin{figure}[!ht]
    \centering
    \includegraphics[width=0.6\textwidth]{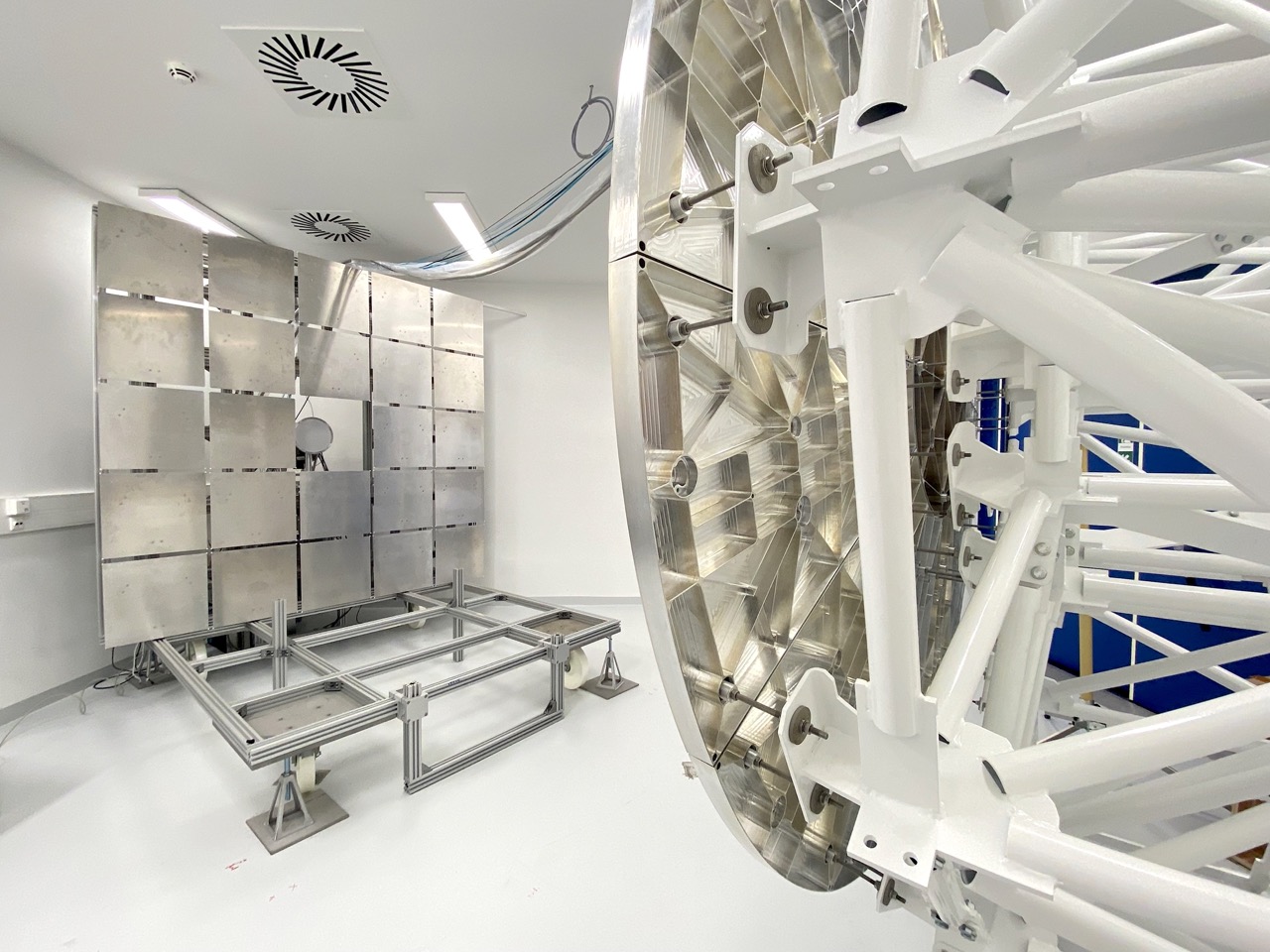}%
    \caption{BRASS-p setup, currently under operation in a radiation-shielded laboratory at University of Hamburg.}  %
    \label{fig:brass-setup}%
\end{figure}

The present prototype setup BRASS-p is shown in Fig.~\ref{fig:brass-setup} and its main parameters are summarised in Table~\ref{table:component_parameter}. BRASS-p is located in a climatised and shielded room with a 60~dB radiation shielding effectiveness up to 18~GHz. The conversion surface of BRASS-p consists of 24 $50\times50$~cm aluminium panels. The 24 panels are arranged in a $5 \times 5$ square pattern, with the central slot left empty for the microwave beam throughput to the receiver, while the gap between the panels is 2 cm. Ultimately, the effective conversion area where the emitted signal arrives to the mirror is $A_0 = 4.44~\text{m}^2$.   

The conversion surface and the parabolic reflector are separated by 4.5 metres and mounted on two separate frames. The frame of the conversion surface also supports a broadband \SIrange{12}{18}{\giga\hertz} receiver located behind the conversion surface. The rest of the detection apparatus is placed outside of the shielded room, in order to reduce the radio frequency interference (RFI) produced by various electronic components.

\begin{table}[!ht]
\centering
\begin{tabular}{c|cc}
\hline
\multirow{3}{*}{Conversion surface}        & \multicolumn{1}{c|}{Panel Size}                  & 50 cm $\times$ 50 cm                 \\ \cline{2-3} 
                                           & \multicolumn{1}{c|}{Number of Panel}             & 24                         \\ \cline{2-3} 
                                           & \multicolumn{1}{c|}{Effective Conversion Area}   & 4.44~m$^2$                      \\ \hline
                                           & \multicolumn{1}{c|}{ Direction}  & 120° (SE)                              \\ \cline{2-3} 
\multirow{3}{*}{Parabolic mirror}          & \multicolumn{1}{c|}{Diameter}                    & 2.5 m                                \\ \cline{2-3}
                                           & \multicolumn{1}{c|}{Focal Length}                & 4.8 m                                \\ \cline{2-3} 
                                           & \multicolumn{1}{c|}{Surface RMS}           & 4-6~$\mu$m                           \\ \hline
\multirow{4}{*}{Receiver frontend} & \multicolumn{1}{c|}{Frequency}                    & 12-18~GHz                            \\ \cline{2-3} 
                                           & \multicolumn{1}{c|}{Polarization}                & Dual Linear (H+V)                                    \\ \cline{2-3} 
                                           & \multicolumn{1}{c|}{Receiver Temperature} & 15 K                                 \\ \cline{2-3} 
                                           & \multicolumn{1}{c|}{Output IF Bandwidth}         & 4096 MHz                                \\ \hline
\multirow{3}{*}{Digital backend}          & \multicolumn{1}{c|}{Maximum Sampling Rate}        & 8192 Msamples/s      \\ \cline{2-3} 
                                           & \multicolumn{1}{c|}{Maximum Output Bandwidth}       & 4096 MHz/polarisation                            \\ \cline{2-3} 
                                           & \multicolumn{1}{c|}{Spectral Resolution}                  & 625~Hz, 125~Hz, 25~Hz\\ \hline
Location                                   & \multicolumn{2}{c}{53°34'40"N,9°53'8"E}                                                \\ \hline
\end{tabular}
\caption{Overview of the BRASS-p key components and their parameters.}         \label{table:component_parameter}
\end{table}

 \subsection{Mechanical and optical setup}\label{mechanical_optical_setup}

 The parabolic reflector of BRASS-p is mounted on a fixed metal frame anchored to the floor. It is assembled from 13 aluminium panels, with each panel having a surface roughness of less than 2~$\mu$m. A laser scan of the entire reflector has shown that its surface deviations from the designed parabolic geometry have an r.m.s of 4~$\mu$m--6~$\mu$m, thus allowing for its to be operated at frequencies of up to $\approx \SI{4}{\tera\hertz}$ for which the standard antenna formalism \citep{1966IEEEP..54..633R} yields a reduction of the gain by 3~dB. 

The other two main optical components of BRASS-p (the conversion surface and the corrugated horn feed of the receiver) are aligned with respect to the parabolic reflector. The receiver feed is oriented along the optical axis of the reflector and placed at its focal point. The conversion surface is orthogonally aligned with the optical axis of the reflector, with the alignment made both for each of the 24 individual panels and for the overall surface. Dedicated ray-tracing calculations have demonstrated that the angular accuracy of $\approx 1^{\circ}$ is required for the alignment to ensure the required optical performance \cite{flehmke}. 

 

The frame of the converter surface is mounted on wheels, and at the first step of the alignment the frame is aligned orthogonally to the optical axis of the reflector. Each of the converter surface panels is then aligned individually with respect to the frame. The orientation of each panel has been verified with two independent
 methods. First, a commercial 3D-laser scanning device (Leica RTC360)
 has been deployed in the middle point of the lab to record the 
 positions of the parabolic reflector as well as of the panels. Fitting a flat profile to the  point cloud of each panel returns the deviation of 0.3 mm. The surface deviation is due to the bending from the mechanical stress between the actuators that attached to the panels. 
 In addition to that, the orientation of each panel was examined using a custom-made alignment set attached to it which comprises three ranging lidars and a 
 CCD camera to register the positions of the lidar spots on the parabolic dish \cite{flehmke}. Geometric analysis of the measurements made with these two approaches
 confirms that the surface normal of the panels and the optical axis
 of the reflector have been aligned within $\approx 0.2^{\circ}$. 

 In order to align the receiver feed, a cap with a laser is mounted on the opening of the feed. The laser traces the optical axis of the feed. The feed is then adjusted so that the laser spot is visible behind the small hole ($2$~mm in diameter) at the centre of the parabolic reflector. With this approach, the optical axes of the receiver feed and the parabolic reflector have been aligned to within $\approx 0.02^\circ$. The offset between the focal point and the phase centre is estimated at $\pm 0.1$~mm.

\subsection{Receiver}\label{sub:receiver}

\begin{figure}[!t]
    \centering
    \includegraphics[width=15cm]{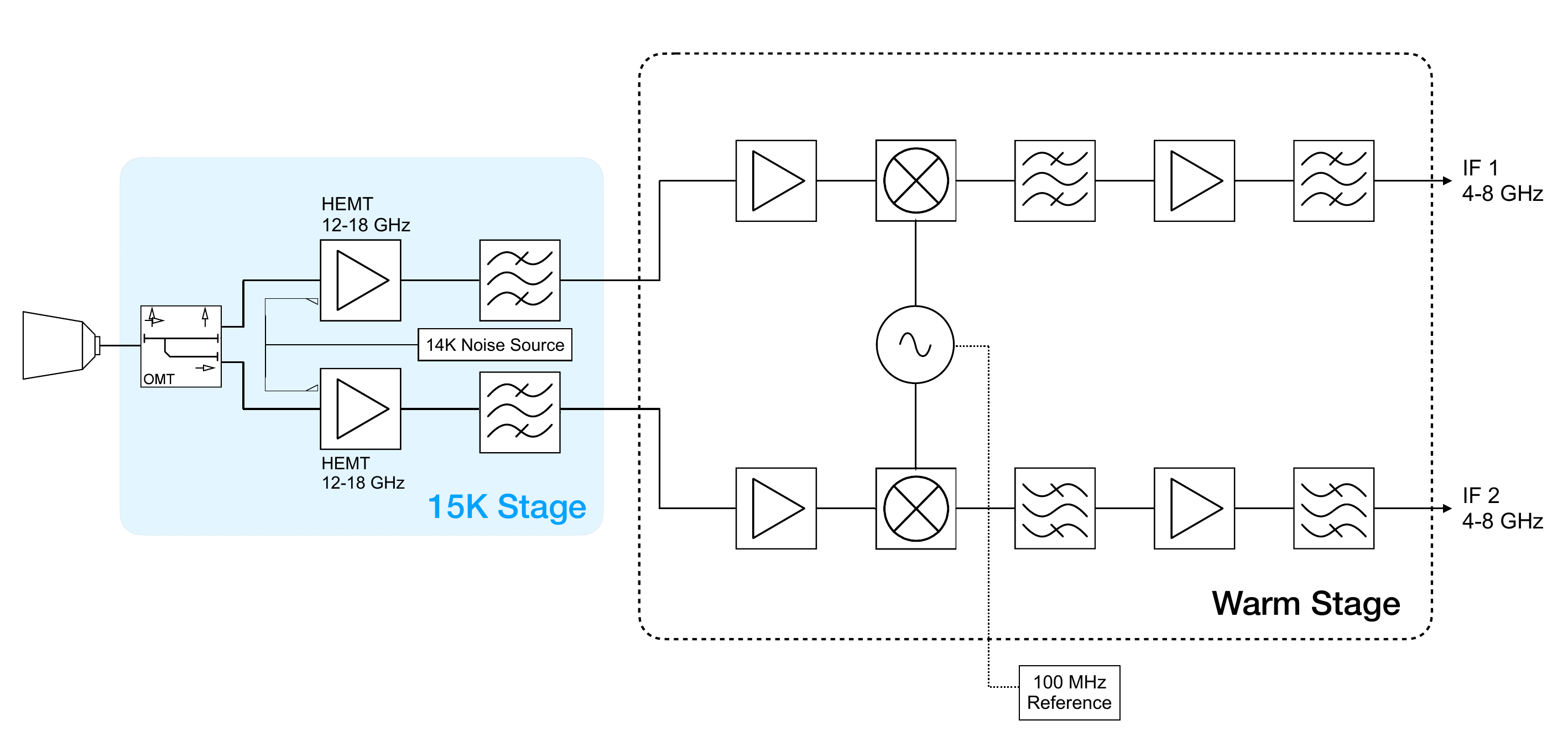}
    \caption{The receiver frontend of BRASS-p.The aperture of the corrugated Ku-band feed is operated at room temperature. The OMT unit separates the two polarisation signals and guides the respective signals to the two LNAs. The LNA is a wideband HEMT amplifier for the \SIrange{12}{18}{\giga\hertz} band, coupled to the \SI{14}{\kelvin} noise source for calibration. The OMT, the noise source, and the LNA are hosted inside a Dewar cryostat maintained at 15~K. The signal is guided outside the cryostat via coaxial cable to the downconverter, providing the 4--\SI{8}{\giga\hertz} IF output. The IF output signal undergoes further gain amplification and bandpass equalisation before being output through an SMA port.}
    \label{fig:analogschematic}
\end{figure}

    BRASS-p operates with a Ku-band receiver designed and manufactured at the MPIfR, Bonn. 
    The receiver covers a frequency range of \SIrange{12}{18}{\giga\hertz}, with horizontal and vertical polarisation channels providing measurements in dual linear polarisation mode. 
    The corrugated shape of the receiver feed ensures a low cross-talk (less than \SI{-40}{\dB}) between the two polarisation channels and high gain over the target frequency. The signal path through the receiver frontend is shown in Fig.~\ref{fig:analogschematic}.
    
    The orthomode transducer (OMT), the calibration noise source, and the high-electron mobility transistor (HEMT) used as low noise amplifier (LNA) are operated at
    a temperature of \SI{15}{\kelvin}.  The noise contributed by these components 
    determines the sensitivity of the receiver.

    The outputs of the LNA are connected via coaxial cables to the 
     first heterodyne mixing stage operated at a room temperature. 
    The heterodyne mixes the signal with a tuneable local oscillator (LO)
    from \SIrange{8}{10}{\giga\hertz}, this LO is synthesised by an external reference signal of 100 MHz. This way, the signal is shifted 
    from the Ku band (\SIrange{12}{18}{\giga\hertz}) to the intermediate frequency (IF) range from \SIrange{4}{8}{\giga\hertz} with a  bandwidth of \SI{4}{\giga\hertz}. 
    The second amplifying stage and equalizer are applied to the IF signal before the second mixing stage and digitizer  integrated in the third generation 
    digital backend system. 

    The receiver performance is characterised using an internal calibration source (noise diode) with a noise temperature of \SI{14}{\kelvin}. This procedure yields $T_\mathrm{rec} = \SI{10}{\kelvin}$ for the average receiver temperature over the full \SIrange{12}{18}{\giga\hertz} frequency range.

    The receiver is controlled by a control computer outfitted with a graphical user interface (GUI) for setting the LO frequency, operating the cryostat valves, and monitoring a number of peripheral parameters of the receiver such as the output power, bias currents, amplifier gains, and operational temperature. 

\subsection{Aperture efficiency}

    Aperture efficiency, $\eta_\mathrm{a}$, of the the BRASS setup describes the effective collecting area of its optical apparatus, $A_\mathrm{eff} = \eta_\mathrm{a} A_0$. The aperture efficiency is a product of individual factors pertaining to the conversion surface, the parabolic reflector, and the feed antenna of the receiver:
    
    \begin{equation}
        \eta_\mathrm{a} = \eta_\mathrm{sf}  \cdot \eta_\mathrm{il}  \cdot \eta_\mathrm{sp} \cdot \eta_\mathrm{opt} \cdot \eta_\mathrm{ph}  \, , \label{eqn:aperture_efficiency}
    \end{equation}
where $\eta_\mathrm{sf}$ is the surface efficiency, $\eta_\mathrm{il}$  and $\eta_\mathrm{sp}$ are the illumination and spillover efficiencies of the receiver feed,  $\eta_\mathrm{opt}$ is the optical diffraction efficiency, and $\eta_\mathrm{ph}$ is the phase centre efficiency. 

\begin{figure}[h!]
        \centering
        \includegraphics[scale =0.6]{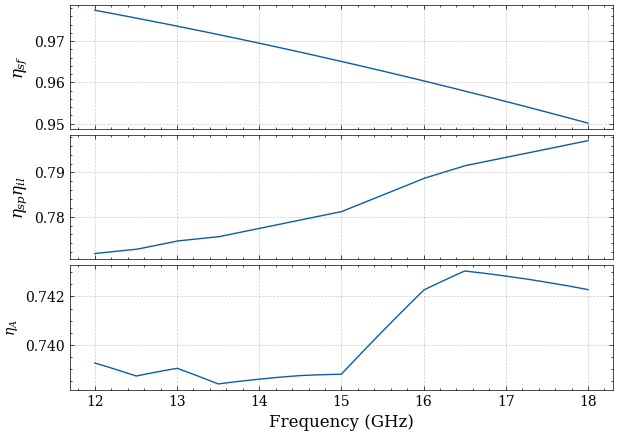}
        \caption{Aperture efficiency, $\eta_\mathrm{a}$ of BRASS-p as described by the equation~\ref{eqn:aperture_efficiency} (bottom panel) and two of its  components: the surface (top panel) and feed illumination and spillover (middle panel) efficiencies.} 
        \label{fig:aperture_efficiency}
    \end{figure}

The surface efficiency, $\eta_\mathrm{sf}$, is the cumulative product of all deviations of the conversion surface and the parabolic reflector from their ideal shapes. The surface rms for both of them is better than \SI{10}{\micro\meter} which does not lead to any degradation of the signal in the \SIrange{12}{18}{\giga\hertz} frequency range. The surface efficiency is therefore effectively determined by the magnitude of bending of the conversion panels ($\sigma_s = \SI{0.3}{\milli\meter}$), and the resulting $\eta_\mathrm{sf}$ ranges from 0.97 to 0.95 in the frequency range of \SIrange{12}{18}{\giga\hertz} (top panel of Fig.~\ref{fig:aperture_efficiency}). 

The product of the illumination efficiency, $\eta_\mathrm{il}$, and the spillover efficiency $\eta_\mathrm{sp}$ of the receiver feed, optimised by a $-$10~dB taper, is shown in the middle panel of Fig.~\ref{fig:aperture_efficiency}. The optical efficiency $\eta_\mathrm{opt} = 0.98$ is estimated from the FEM simulation described in Section~\ref{sec:optical_scheme}.

The phase centre offset efficiency $\eta_\mathrm{ph}$ describes the effect of an offset between the phase centre of the receiver horn and the focal point of the parabolic reflector. Measurements made with a laser scanner show this offset to not exceed $\pm 0.1~\text{mm}$, thus not introducing any signal degradation \citep{Ingerson_Rusch1973} and allowing for $\eta_\mathrm{ph}=1$ to be applied. 

The resulting total aperture efficiency, shown in the bottom panel of Fig.~\ref{fig:aperture_efficiency}, is $\eta_\mathrm{a}\approx 0.72$ and varies only weakly over the \SIrange{12}{18}{\giga\hertz} frequency range. With this efficiency BRASS-p has an effective collecting area $A_\mathrm{eff} = 3.22~\mathrm{m}^2$.


\subsection{Data acquisition system} \label{sub:digitizer}

\begin{figure}[ht!]
    \centering
    \includegraphics[width=15cm]{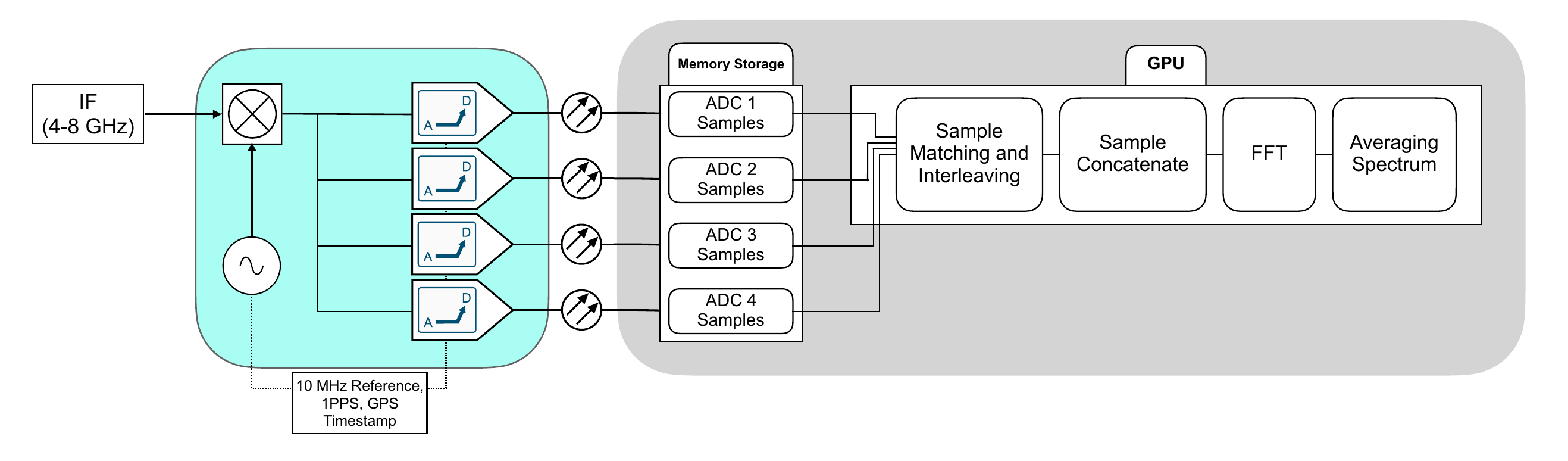}
    \caption{Signal processing path for a single polarisation channel of BRASS-p. The 4--8~GHz IF signal from the receiver frontend is first downconverted to the 0--4~GHz band and then split into four streams digitised separately in the respective analog-to-digital converter (ADC) modules. Each ADC module operates at 2048 Megasamples/s (Msps) and is phase shifted by $\frac{\pi}{2}$ with respect to the adjacent ADC modules. The digitally sampled output of the ADC is packed, timestamped, and sent to the acquisition computer via four fibre links. The samples are placed in a temporary memory storage from which they are accesed by a GPU-based program that matches the timestamps and interleaves the samples to obtain the true time series data at 8192 Msps. The interleaved time series is then concatenated and sent to the fast Fourier transform (FFT) module. The resulting power spectrum is stacked and saved to disk.}
    \label{fig:digitalschematic}
\end{figure}

For each of the two polarisation channels, the IF signal delivered by the receiver is processed as shown in Fig.~\ref{fig:digitalschematic}. The signal is first digitised in a digital baseband converter unit DBBC3 \cite{tuccariDBBC3VLBI322012} designed and manufactured for BRASS-p by HAT-Lab (www.hat-lab.cloud) and MPIfR, Bonn. The DBBC3 unit digitises up to eight input signal streams simultaneously and delivers an instantaneous bandwidth of 4 GHz per polarisation. In this process, the two IF signals from the two  polarisation channels are downconverted to the 0--4 GHz band and digitised by four interleaved analogue-to-digital converter (ADC) modules, each operating at 2048 megasamples/s (Msps). In order to optimize the gain stages in each ADC, each ADC's power is measured, and a correction factor is applied in dB fractions to get from each of the ADCs a very similar value with a precision of less than 1 \%. A novel technique was employed to finely adjust delay sampling between consecutive ADCs  by utilizing the digital control capabilities of the device's input clock delay, synchronizing the sampling time of two adjacent ADCs, and followed by the cross-correlation of the obtained data. Maximizing this correlation provides precise information for setting the delay control to achieve the desired final delay. With this method, the delay time error is down to a fraction of the picosecond, thus minimizing sampling artefacts. 

During BRASS-p data taking runs, the digitisation was performed
over a fixed integration time and the digitised output was stored and processed in a dedicated computer. 

The data processing comprised interleaving and concatenation of the input streams, followed by a fast Fourier transform (FFT) of the concatenated data stream and subsequent storing and averaging of the resulting power spectra. The FFT processing was parallelised using a graphical computing unit (GPU) with high performance shader modules coupled with video random access memory (VRAM). The parallelisation used the algorithm developed for the WISPDMX experiment \cite{Nguyen:2019xuh}. This allows the FFT to operate at a time efficiency of 60\% when providing 160 million spectral channels (frequency resolution of \SI{25}{\hertz}) across the \SI{4}{\giga\hertz} bandwidth. For the BRASS-p \SIrange{12}{18}{\giga\hertz} frequency band, this corresponds to ${\Delta \nu}/{\nu} =(1.4$--$2.1)\times 10^{-9}$, which is fully sufficient for addressing both the standard halo scenario and the DM stream models. Further acceleration of the data processing can be achieved by parallelising the FFT with the data sample acquisition, which can be implemented after an upgrade of the data acquisition computer. Presently, a single data acquisition segment of BRASS-p lasts 19.8 seconds. The data stream was arranged into 2475 fragments with 65536000 samples each. At the sampling rate of 8192 Msps, the respective frequency resolution is \SI{125}{\hertz}.

\subsection{Measurements}\label{sec:measurement}

    Data taking runs of BRASS-p consist of multiple data taking segments described above. These segments are set up before a given data taking run and then executed automatically. First, the receiver frontend is configured, using its GUI interface to control the peripheral parameters of the receiver. The frontend GUI is subsequently used for monitoring receiver performance during the data taking run. Next, the DBBC3 (cyan block in Fig. \ref{fig:digitalschematic}) is set up, using its client terminal interface which controls and monitors the signal gain and the output sample distribution. At the final step, the control computer is engaged to run a script that manages the data taking run, including the collection and processing of the DBBC3 output by two desktop computers, the FFT processing of the digitised time series data, and the subsequent storage of the processed spectra.

\begin{figure}[ht!]
        \centering
        \includegraphics[scale =0.6]{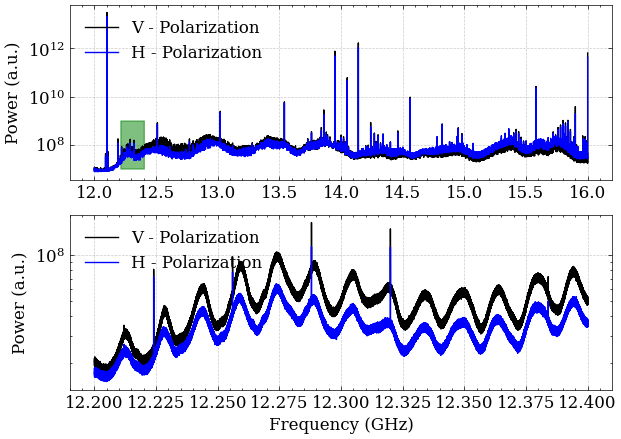}
        \caption{Examples of output spectra for the two polarisations measured by BRASS-p in a single \SI{19.8}{\second} segment. The top panel shows the full \SI{4}{\giga\hertz} spectra and the bottom panel shows a zoom over a \SI{200}{\mega\hertz} part of the spectra. Narrow band spikes seen in the spectra result from interference with various electronic components of the setup, and these spikes will have to be excised during the detailed search for a DM signal. The oscillatory pattern visible in the \SI{200}{\mega\hertz} fragment of the spectra is produced by standing waves due to multiple reflections between the main optical components of the setup.} 
        \label{fig:single_spectrum_2pol}
    \end{figure}

    Fig.~\ref{fig:single_spectrum_2pol} shows an example of raw power spectra obtained from a single \SI{19.8}{\second} segment of BRASS measurements. The narrow-band spikes seen in the spectra result from interference introduced by various electronic components of the setup. 
    Most of these peaks arise from the interference of the \SI{10}{\mega\hertz}  and have a width of exactly one spectral channel (\SI{125}{\hertz}).
    These spikes will normally be excised during the dedicated signal searches. The oscillatory pattern visible in Fig.~\ref{fig:single_spectrum_2pol} in the \SI{200}{\mega\hertz} zoom in the spectra results from standing waves produced through multiple reflections between different optical components of the setup. This pattern can be partially corrected during the system noise calibration of the setup.
    
\subsection{Total power and system noise calibration}
\label{Section:Calibration}

\begin{figure}[ht!]
        \centering
        \includegraphics[scale =0.6]{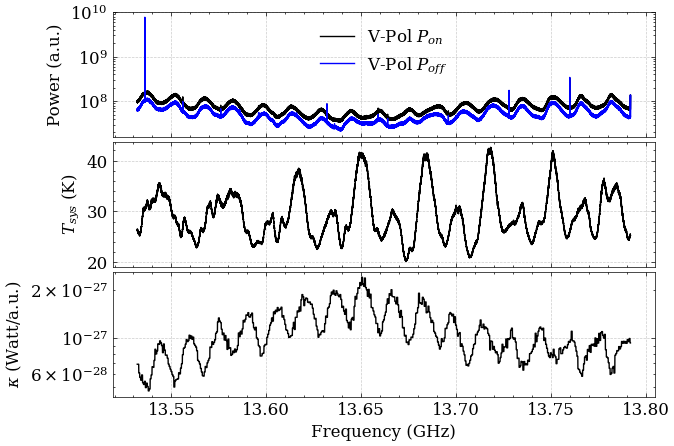}
        \caption{Total power calibration procedure using the on/off measurements with the noise diode, illustrated with a \SI{240}{\mega\hertz} fragment of the spectral measurements. Shown are the on/off power spectra, $P_\mathrm{on}(\nu)$ and $P_\mathrm{off}(\nu)$,  measured in the V-polarisation channel (top panel) and the resulting system temperature, $T_\mathrm{sys}(\nu)$ (middle panel) and calibration factor, $\kappa(\nu)$ (bottom panel).   }  
        \label{fig:T_sys_conversion_factor}
    \end{figure}

The total power and system noise calibration of BRASS-p measurements is carried out by measuring the frequency-dependent system temperature $T_\mathrm{sys}(\nu)$ of the setup from on/off spectra obtained with the $T_\mathrm{cal}$ signal of the noise diode, which yields
\begin{equation}
    T_\mathrm{sys}(\nu) = T_\mathrm{cal} \left(\frac{P_\mathrm{on}(\nu)}{P_\mathrm{off}(\nu)}-1\right)^{-1} \,, \label{eqn:tsys}
\end{equation}
where $P_\mathrm{on}$ and $P_\mathrm{off}$ are the power spectra measured with and without the $T_\mathrm{cal}$ signal added. The total power calibration factor, $\kappa(\nu)$ is then given by
\begin{equation}
    \kappa(\nu) = \frac{k_\mathrm{B}\, T_\mathrm{sys}(\nu)\, \Delta\nu^{1/2}\, \tau^{-1/2}}{\sigma_\mathrm{off}(\nu)} \,,
\end{equation}
where $\sigma_\mathrm{off}(\nu)$ is the standard deviation (or noise power) in the raw spectrum. The calibration procedure is illustrated in Fig.~\ref{fig:T_sys_conversion_factor} showing a \SI{240}{\mega\hertz} fragment of the on/off spectra and the resulting $T_\mathrm{sys}(\nu)$ and $\kappa(\nu)$ obtained from these spectra. This calibration procedure also provides at least partial correction for the standing waves, since the $T_\mathrm{cal}$ signal from the noise diode undergoes the same reflections as the incoming radiation generated at the conversion surface. This can be observed in the regular patterns also appearing in the $T_\mathrm{sys}(\nu)$ and $\kappa(\nu)$ plotted in Fig.~\ref{fig:T_sys_conversion_factor}. 

    \begin{figure}[ht!]
        \centering
        \setlength{\unitlength}{0.05\textwidth}
        \begin{picture}(15,7)
         \put(-2.7,0){\includegraphics[scale =0.5]{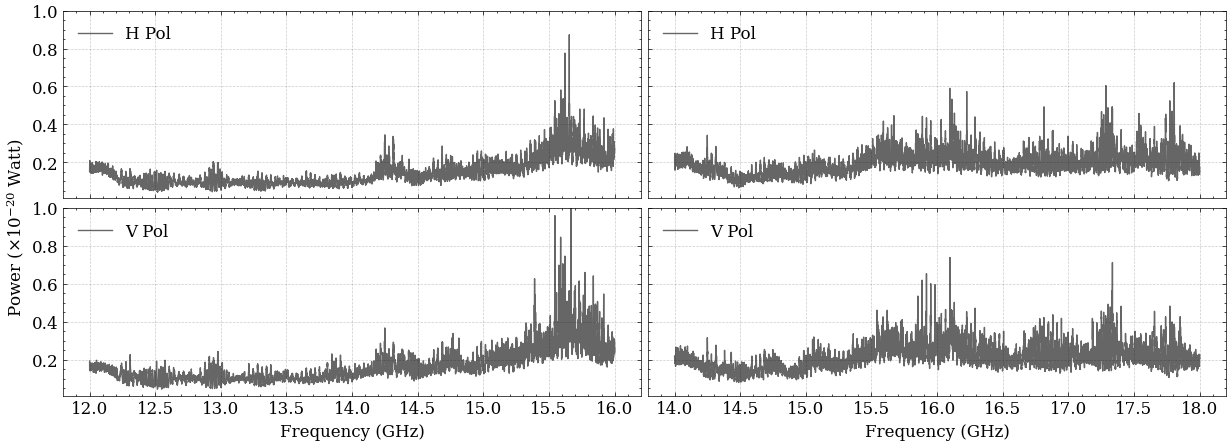}}
         \end{picture}
        \caption{Calibrated noise power per spectral channel, $\sigma_\mathrm{det}(\nu)$,  for the frequency range of \SIrange{12}{16}{\giga\hertz}, obtained for the vertical (V) and horizontal (H) polarisation channels from a single \SI{19.8}{\second} measurement of BRASS-p. The narrow band spikes discussed in section \ref{sec:measurement} have been removed before the calibration. In each polarisation channel, a detection sensitivity of $\sigma_\mathrm{det} \approx \SI{1e-21}{\watt}$ is reached.} 
        \label{fig:cal_VH_spectrum}
    \end{figure}

The average $T_\mathrm{sys}$ obtained from the on/off calibration is about \SI{30}{\kelvin}, which is close to what can be expected after accounting for the major factors contributing to it. These factors are the receiver noise temperature, $T_\mathrm{rec}\approx \SI{10}{\kelvin}$, the effective antenna temperature $T_\mathrm{ant} = (1 - {\cal R}) \,T_\mathrm{room} \lesssim \SI{3}{\kelvin}$ (for the antenna reflectivity ${\cal R}=0.99$ and the room temperature $T_\mathrm{room} =\SI{20}{\celsius}$), and the spillover temperature, $T_\mathrm{spill} = \epsilon_\mathrm{f}\, T_\mathrm{room} \approx \SI{11}{\kelvin}$, of thermal radiation from the room also entering the receiver feed, for which $\epsilon_\mathrm{f} = P_\mathrm{lobe}/P_\mathrm{beam} \approx 0.04$ can be estimated. Summed up, these three factors yield a temperature of \SI{24}{\kelvin}, and we expect that the remaining difference of $\approx \SI{6}{\kelvin}$ between this value and the average measured $T_\mathrm{sys}$ can result from the multiple reflections responsible for the standing waves.

The respective calibrated noise spectra, $\sigma_\mathrm{det}(\nu)$, obtained from a single \SI{19.8}{\second} segment are shown in Fig. \ref{fig:cal_VH_spectrum} for the \SIrange{12}{16}{\giga\hertz} band, indicating that the lowest detectable power of $\sigma_\mathrm{det} \approx \SI{e-21}{\watt}$ can be achieved by BRASS-p in such a measurement.
    
\begin{figure}[ht!]
        \centering
        \setlength{\unitlength}{0.05\textwidth}
        \begin{picture}(15,7)
            \put(-2.7,0){\includegraphics[scale =0.5]{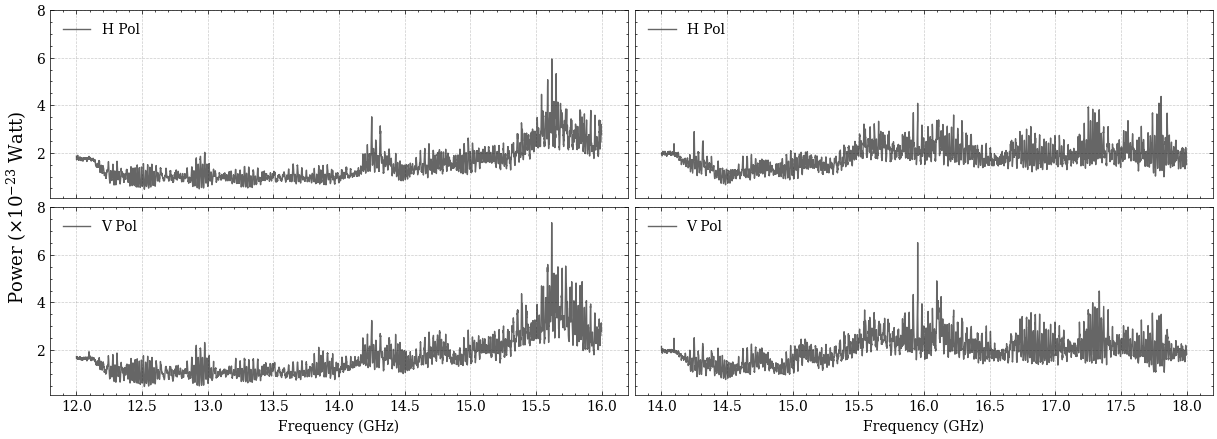}}
        \end{picture}
        \caption{Stacked noise power per spectral channel, $\sigma_\mathrm{det}(\nu)$, obtained from BRASS-p first science measurement runs SR1a (left panels) and SR1b (right panels) with the total measurement times of \SI{59.99}{\hour} and \SI{66.07}{\hour}, respectively. A detection sensitivity of $\sigma_\mathrm{det} \approx 10^{-23}$~W is reached in these measurements. } 
        \label{fig:lowest_det_pow_VH}
    \end{figure}

\subsection{Results from the first science run of BRASS-p}
\label{Section:Results}

The first science run (SR1) of BRASS-p was carried out in November and December 2022, and it consisted of two data taking periods (SR1a and SR1b) in which the BRASS-p setup was used for making respective measurements in the frequency ranges of \SIrange{12}{16}{\giga\hertz} and \SIrange{14}{18}{\giga\hertz} frequency ranges. Table~\ref{table:rundownSR1} provides a summary of these runs.

    \begin{table}[H]
        \centering
        \setlength{\tabcolsep}{12pt}
        \begin{tabular}{rcc}
                                  &                                         \textbf{SR1a}  & \textbf{SR1b}   \\ \hline
            Frequency band:    &                                         12-16 GHz      & 14-18 GHz       \\
            Number of polarization channels:    &                                          2     &   2       \\
            Spectral resolution:           & 125 Hz & 125 Hz                                                   \\
             Starting date:           &                                          13 Nov 2022  &   24 Nov 2022 \\
            Ending date:        &                                          19 Nov 2022  &   2 Dec 2022 \\
            Number of spectra collected: &                                              9998           &12012           \\ 
            Total measurement time: &                                              54.989 hours         &   66.066 hours           \\ \hline
        \end{tabular}
        \caption{Summary of BRASS-p measurement runs SR1a and SR1b}         \label{table:rundownSR1}
    \end{table}

For each of the data taking runs, the individual spectra have been averaged, yielding cumulative noise spectra. These spectra are plotted in Fig.~\ref{fig:lowest_det_pow_VH} separately for the two polarisation channels. As expected, the cumulative spectra reach a noise level of $\sigma_\mathrm{det} \approx 10^{-23}$~W, signifying a factor of $\approx 100$ improvement over the noise measured in single \SI{19.8}{\second} data acquisitions. 

First inspections of the spectra obtained do not reveal an obvious signal candidate, which follows the Maxwellian distribution of the dark matter particle's velocity. Deferring detailed signal searches to a later publication, we apply here the cumulative noise spectra for calculating the effective detection sensitivity of the BRASS-p SR1 measurements for the kinetic mixing, $\chi(\nu)$, of hidden photons. Applying the measured noise power per spectral channel, $\sigma_\mathrm{det}(\nu)$, to the equation~\ref{eq:power_dish}, and taking into account that a virialized DM signal should be distributed over $N_\mathrm{ch}(\nu)$ channels \citep{nguyenFirstResultsWISPDMX2019}, we obtain
\begin{equation}
        \chi(\nu) = 3.2 \times 10^{-14} \left( \frac{\sigma_\mathrm{det}(\nu) N_\mathrm{ch}(\nu)^{-1/2}}{ 10^{-23} \text{ W}}   \right)^\frac{1}{2} \left( \frac{0.3 \text{ GeV}/\text{cm}^3}{\rho_\text{DM}} \right)^\frac{1}{2} \left( \frac{1 \text{ m}^2}{A_\mathrm{eff}}\right)^\frac{1}{2} \left(\frac{\sqrt{2/3}}{\sqrt{\alpha}} \right)\,,
\label{eq:sensitivity}
\end{equation}
where $\alpha = 1/3$ for a single polarisation channel and $\alpha = 2/3$ for the two polarisation channels combined. 

\begin{figure}[h!]
    \centering
    \includegraphics[scale =0.6]{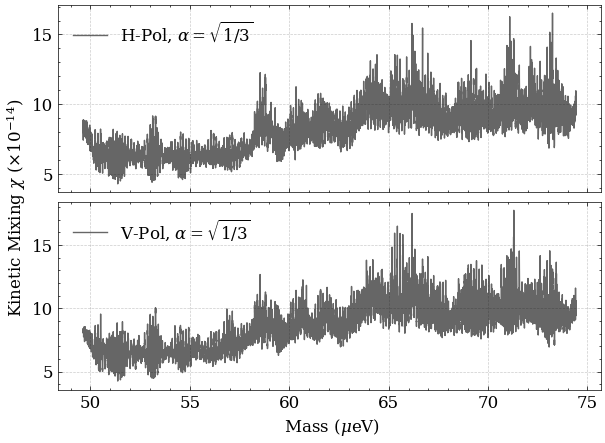}
    \caption{Combined 1-$\sigma$ detection sensitivity of BRASS-p measurements plotted for each polarisation channel over the entire range of the hidden photon mass probed by the SR1 measurements.} 
    \label{fig:sensitivity}
\end{figure}

The resulting $\chi(\nu)$ are combined for the entire range of the hidden photon mass probed
by the SR1 measurements of BRASS-p and plotted in Fig.~\ref{fig:sensitivity} for each polarisation channel. Similar sensitivities are achieved in both polarisation channels, benchmarking on $\chi \lesssim 10^{-13}$ over the full range of measurements. Further improvement by a factor of $\approx \sqrt{2}$ can be achieved by combining the two polarisation channels together, and this will be done together with the planned detailed signal search.

\begin{figure}[ht!]
    \centering
    \includegraphics[scale =0.8]{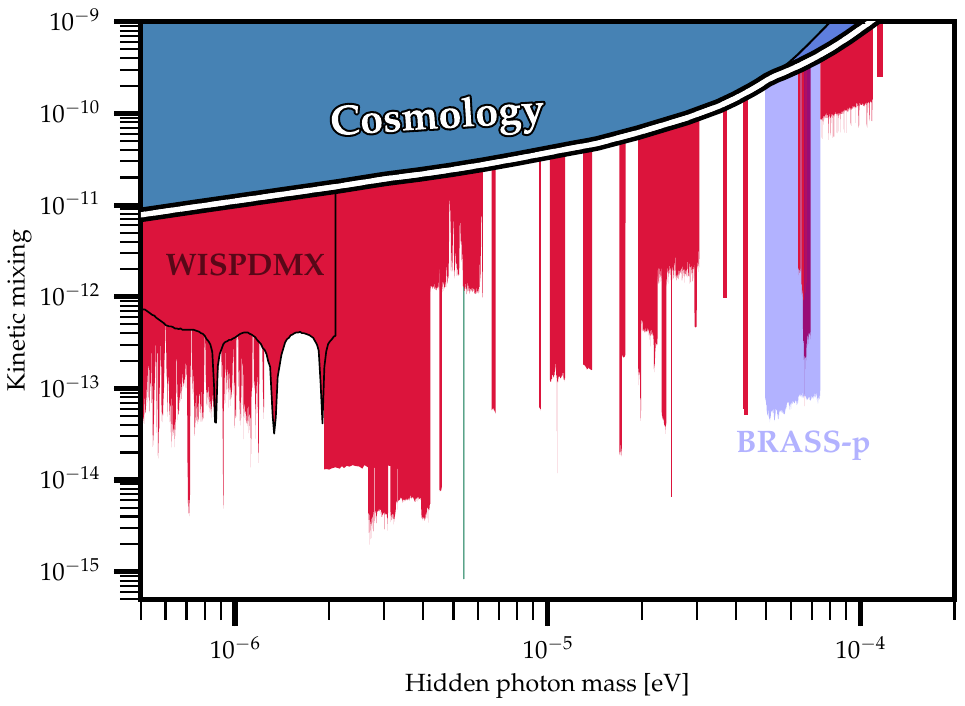}
    \caption{Detection sensitivity of the BRASS-p SR1 measurements in the vertical polarisation channel, compared to exclusion limits obtained in other hidden photon detection experiments (see \citep{AxionLimits}).}
    \label{fig:hplimits}
\end{figure}

To put the BRASS-p SR1 measurements in the broader context of hidden photon searches, the detection sensitivity obtained for the vertical polarisation channel is compared in Fig.~\ref{fig:hplimits} to exclusion limits obtained in other experiments. This comparison shows that BRASS-p reaches the best sensitivity to the kinetic mixing of hidden photon dark matter over the entire range of particle mass probed by its SR1 measurements.

\newpage
\section{Conclusions}

This paper, and the BRASS-p experiment described in it, demonstrate the overall feasibility of searches for dark matter signals generated at a conducting {\em converter} surface \citep{hornsSearchingWISPyCold2013} to be employed for broadband radiometric searches for hidden photon and axion DM in the radio-to-submillimetre regime, from $\sim$\SI{1}{\giga\hertz} to $\sim$\SI{1}{\tera\hertz}, in which heterodyne receivers can be effectively applied for detecting signals over broad bands and at a spectral resolution $\Delta\nu/\nu \lesssim 10^{-6}$. The BRASS-p setup deals successfully with all the critical aspects of generating and detecting an electromagnetic signal from dark matter particles passing through the converter surface, and the first BRASS-p measurements reach the power sensitivity of $10^{-23}$~W in the \SIrange{12}{18}{\giga\hertz} frequency range, which translate into the world record sensitivity of $\chi < 10^{-13}$ to the kinetic mixing of DM hidden photons.

The initial results from BRASS-p measurements presented in this paper will be further refined through a dedicated signal search which would adapt and expand the methodology developed for such searches in the broadband part of the haloscope experiment WISPDMX \citep{nguyenFirstResultsWISPDMX2019}. The data acquisition will be updated to include a digital FX (DiFX) \citep{2011PASP..123..275D} correlation of the two polarisation channels, thus enabling the full polarisation reconstruction and an effective feed rotation \citep{2021hai1.book..127R} for probing all possible electric vector orientations of the signal and hence expanding substantially the capabilities of HP dark matter searches with BRASS-p. Multiple epoch observations would further explore seasonal effects expected due to the Earth orbital motion \citep{2021PhRvD.104i5029C} as well as the possibility for caustic focusing of the dark matter signal \citep{PhysRevD.105.063032}. 

\begin{figure}[ht!]
    \centering
    \includegraphics[scale =0.8]{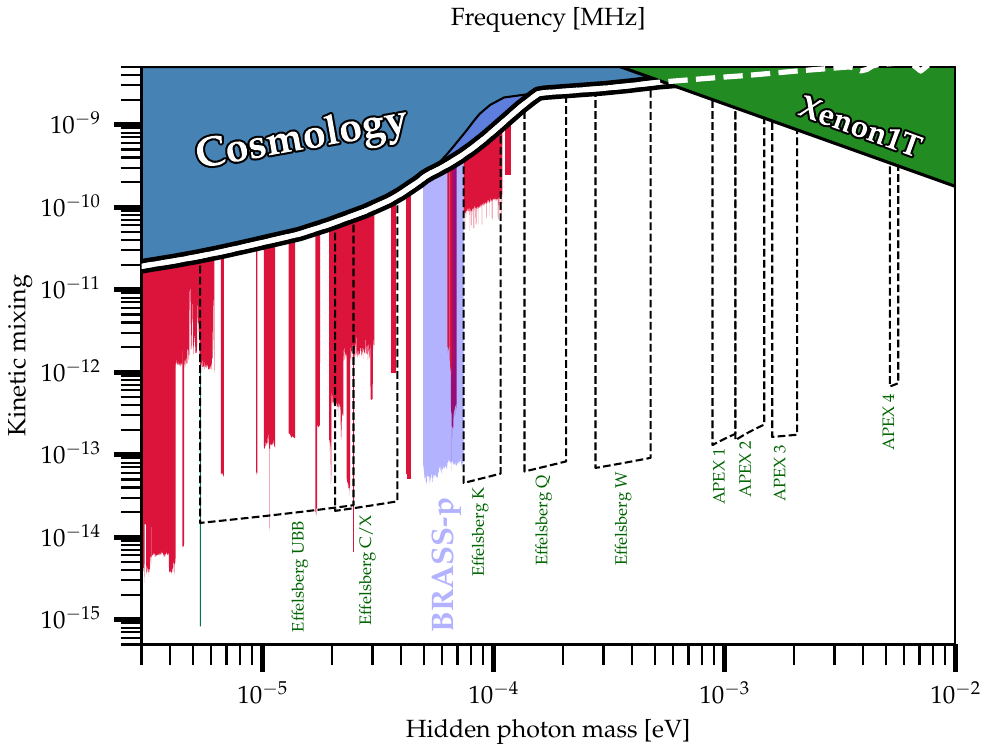}
    \caption{BRASS-p SR1 measurements complemented with the expected detection sensitivities (shown with dashed lines) which can be achieved in 10-day measurement runs using the existing BRASS-p setup with one of nine broadband receivers constructed for the Effelsberg 100-metre telescope and the 15-metre APEX telescope.}
    \label{fig:hpexpectations}
\end{figure}

Hidden photon DM searches with BRASS-p can now be readily extended over a number of bands in the \SI{1}{\giga\hertz}--\SI{1}{\tera\hertz} frequency range, thanks to excellent synergies with the technologies already available and under current development for radio astronomical measurements. Presently, nine different receivers constructed for the Effelsberg and APEX telescopes can be reproduced and used with the existing BRASS-p setup, delivering excellent sensitivity to the hidden photon dark matter. Fig.~\ref{fig:hpexpectations} shows the expected kinetic mixing sensitivity that can be achieved by BRASS-p with these receivers, assuming a 10-day measurement run for each of them. A number of novel ultra-wideband receivers including BRAND \citep{2023ivs..conf...72T} and the ALMA receiver set \citep{2019athb.rept.....R} will allow for an even wider exploration of the frequency/particle mass parameter space.

BRASS measurements will benefit strongly from further synergies with radio astronomical development in the area of data acquisition, and in particular from the DBBC4 developments \citep{2023ivs..conf...72T} which would provide the possibility to increase the recording bandwidth to \SI{64}{\giga\hertz}, thus allowing to cover most of the receiver bands within a single measurement run.

Outfitted with permanently magnetised panels composed of NdFeB magnets \citep{2002JMMM..248..432B} providing an $\approx$\SI{1}{\tesla} magnetic field over a $\approx$\SI{1}{\centi\meter} scale height, BRASS-p will also be used for carrying out searches for the ALP dark matter at frequencies above $\approx$\SI{10}{\giga\hertz}. BRASS-p will not have a sufficient sensitivity to probe the QCD axion band. The latter would require making a 100 day measurement run with an instrument for which the product $B^2\,A_\mathrm{eff}$, of the squared magnetic field strength and the effective collecting area should reach $\approx 100~\mathrm{T}^2\,\mathrm{m}^2$. BRASS-p will be operating at $B^2\,A_\mathrm{eff} \approx 3~\mathrm{T}^2\,\mathrm{m}^2$ thus requiring an about 30-fold increase of this figure of merit for targeting the axion dark matter. This increase will be the main aim of further BRASS developments, and this aim can be achieved either by enlarging the collecting area of the setup (potentially also employing multiple chambers for that) or employing stronger permanent magnets which may become available.

Based on the discussion presented in this paper, we can conclude that the successful demonstration of feasibility of broadband DM searches provided by BRASS-p establishes it as a robust experimental facility for hidden photon DM searches in the \SI{1}{\giga\hertz} to \SI{1}{\tera\hertz} frequency range and builds a strong foundation for further developments of this approach and its ultimate transformation into a long-term facility for very efficient HP, ALP, and axion dark matter searches.

\newpage
\acknowledgments
We thank Marjolein Verkouter, Uwe Bach, Jan Wagner, and Oliver Polch for their useful technical support and Alan Roy for useful comments and suggestions for the manuscript. DH, LHN, and MT acknowledge the support by the Deutsche Forschungsgemeinschaft (DFG, German Research Foundation) under Germany’s Excellence Strategy – EXC 2121 „Quantum Universe“ – 390833306. 
\bibliographystyle{JHEP}
\bibliography{Reference.bib}
\end{document}